\documentclass[aps,prx,twocolumn,superscriptaddress,showpacs]{revtex4-1}

\usepackage{amsmath,amssymb,graphics,epsfig,epstopdf,color,verbatim,ulem,braket,tabularx}
\usepackage[colorlinks,linkcolor=blue,citecolor=blue,urlcolor=blue]{hyperref}

\graphicspath{{Figures/}}

\begin{document}

\title{Topological invariants for interacting topological insulators: II.  Breakdown of the Green's function formalism}
\author{Yuan-Yao He}
\author{Han-Qing Wu}
\affiliation{Department of Physics, Renmin University of China, Beijing 100872, China}
\author{Zi Yang Meng}
\affiliation{Beijing National Laboratory for Condensed Matter Physics, and Institute of Physics, Chinese Academy of Sciences, Beijing 100190, China}
\author{Zhong-Yi Lu}
\affiliation{Department of Physics, Renmin University of China, Beijing 100872, China}

\begin{abstract}
Topological phase transitions in free fermion systems can be characterized by the closing of single-particle gap and the change in topological invariants. However, in the presence of electronic interactions, topological phase transitions can be more complicated. In paper I of this series (arXiv:1510.07816), we have proposed an efficient scheme to evaluate the topological invariants based on the single-particle Green's function formalism. Here, in paper II, we demonstrate several interaction-driven topological phase transitions (TPTs) in two-dimensional (2D) interacting topological insulators (TIs) via large-scale quantum Monte Carlo (QMC) simulations, based on the scheme of evaluating topological invariants presented in paper I. Across these transitions, the defining symmetries of the TIs have been neither explicitly nor spontaneously broken. In the first two models, the topological invariants calculated from the Green's function formalism succeed in characterizing the topologically distinct phases and identifying interaction-driven TPTs. However, in the other two models, we find that the single-particle gap does not close and the topological invariants constructed from the single-particle Green's function acquire no change across the TPTs. Unexpected breakdown of the Green's function formalism in constructing the topological invariants is thus discovered. We thence classify the topological phase transitions in interacting TIs into two categories in practical computation: those that have noninteracting correspondence can be characterized successfully by the topological invariants constructed from the Green's functions, while for the others that do not have noninteracting correspondence, the Green's function formalism experiences a breakdown but more interesting and exciting phenomena, such as emergent collective critical modes at the transition, arise. Discussion on the success and breakdown of topological invariants constructed from the Green's function formalism in the context of symmetry protected topological (SPT) states is presented.
\end{abstract}


\date{\today} \maketitle

\section{INTRODUCTION}
\label{sec:Introduction}
This is paper II of the series on "topological invariants for interacting topological insulators". In paper I of this series~\cite{He2015b}, we have proposed an efficient scheme to evaluate the topological invariants based on single-particle Green's function formalism. By introducing a ${\it periodization}$ scheme we have successfully overcome the ubiquitous finite-size effect of topological invariants in QMC simulations of interacting TIs and obtained ideally quantized spin Chern number in one-body-parameter-driven topological phase transition between TIs (or between TI and topologically trivial insulators). In this paper, we apply the numerical evaluation scheme developed in paper I to wider classes of interacting TIs, where across the interaction-driven topological phase transitions, the defining symmetries of the topological insulators have been neither explicitly nor spontaneously broken, but the topological invariants constructed from the Green's function may experience unexpected breakdown. Thus, even though there are cases where the interaction-driven TPTs can be successfully captured by the Green's function scheme, we found that in several TPTs the topological invariants fail and provide artificial information about the transition, as well as the phases. We attribute such difference to the fact that in cases where the Green's function formalism works, the phases across the interaction-driven TPT have noninteracting correspondence, but in the cases where the Green's function formalism fails, there is no noninteracting correspondence to the phases after the interaction-driven TPT. Consequently, calling for more complete understanding of topological invariants in interacting TIs is clearly manifested.

As discussed in paper I~\cite{He2015b}, for noninteracting TIs, both $Z_2$ invariant~\cite{Kane2005b,Fu2007b} and spin Chern number~\cite{Prodan2009} can be simply evaluated from Hamiltonian matrix. For example, in systems with spatial inversion symmetry, the $Z_2$ invariant can be calculated as products of parity eigenvalues of all occupied bands at time-reversal invariant (TRI) points in Brillouin zone (BZ)~\cite{Fu2007b}. Spin Chern number can be calculated by integrating the Berry curvature over BZ~\cite{volovik2009universe}. For interacting TIs, numerical evaluations of the topological invariants become more involved. The generalizations of constructing topological invariants from single-particle Green's function~\cite{Avron1983,So1985,Ishikawa,Volovik1988,Wang2010,Gurarie2011,volovik2009universe} and twisted boundary phases~\cite{Niu1985,Wang2014} have been proposed respectively, whereas the former and in particular its zero-frequency version, has been systematically developed in Ref.~\onlinecite{Wang2010,Wang2011,Wang2012a,Wang2012b,Wang2012c}. There are successful applications of the Green's function formalism in one-dimensional (1D) Su-Schrieffer-Heeger model~\cite{Yoshida2014}, in two-dimensional (2D) Kane-Mele-Hubbard model with various generalizations~\cite{Meng2014,Hung2013,Hung2014,Chen2015,Grandi2015,He2015b}, in Bernevig-Hughes-Zhang model~\cite{Yoshida2012,Budich2013,Amaricci2015}, and also in real material calculations with LDA+Gutzwiller and LDA+DMFT where SmB$_{6}$~\cite{Lu2013} and PuB$_{6}$~\cite{Deng2013} have been predicted to be realizations of correlated TIs from calculating $Z_2$ invariant.

However, this is just the tip of iceberg. The interplay between topology and electronic interaction is expected to lead to more complicated and richer physics. Many possible generalizations of the concept of TI to interacting systems have been put forward. Many exotic phenomena of interacting TIs have been predicted/discovered, such as topological Kondo insulator~\cite{Dzero2010,Jiang2013,Lu2013,Deng2013,Xu2014}, topological Mott insulator and fractionalized TI in 2D and 3D systems~\cite{Pesin2010,Maciejko2015}, interaction-reduced classification of noninteracting TIs in the 10 distinct classes (the tenfold way~\cite{Kitaev_Class,Ludwig_Class1,Ludwig_Class2}) in 1D, 2D, and 3D systems~\cite{Fidkowski2010,Fidkowski2011,Shinsei2012,Lu2012,XLQi2013,HYao2013,Fidkowski2013,WangChong2014b,ZCGu2014,YZYou2014_TSC,YZYou2014_SPT,
YZYou2014_ITI,Morimoto2015}, and interaction-driven intrinsic topological order at the boundary of TIs~\cite{Fidkowski2013,WangChong2013a,WangChong2013b,Bonderson2013,Fidkowski2014,WangChong2014a,Metlitski2015a,WangChong2015,Metlitski2015b}. Besides fermionic systems, it was also proposed that bosonic systems can also form exotic states that are similar to fermionic TIs~\cite{Chen2012,Lu2012,Chen2013,WangChong2013a,Liu2013}, all of which are generally called symmetry protected topological (SPT) states.

More recently, an exotic interaction-driven TPT in 2D system (the BKMH-$J$ model below) that is fundamentally different from the TI-to-trivial insulator transition in noninteracting systems, was discovered with large-scale QMC simulations~\cite{Slagle2015,He2015a}. Across the transition, fermions never close their gap, but emergent collective bosonic modes become critical. Thus one can view this transition as a transition between a bosonic SPT state and a trivial featureless Mott insulator~\cite{He2015a,You2015}. Within appropriate parameter region, this transition is described by a $(2 + 1)$D $O(4)$ nonlinear sigma model with exact $SO(4)$ symmetry, and a topological term at exactly $\Theta=\pi$. It was proposed that, there is a series of such interaction-driven TPT in 2D interacting TIs~\cite{You2015}, and they can be studied with unbiased QMC simulations without minus-sign problem (the BKMH-$J$ and BKMH-$V$ models discussed here are among them). A natural question then arises, namely, what will be the fate of the topological invariants constructed from single-particle Green's function formalism across these exotic interaction-driven TPTs. Here, in this paper, we provide the answer.

In this work, employing large-scale QMC simulations, we explore several interaction-driven TPTs in 2D TIs with neither explicit nor spontaneous symmetry breaking. These TPTs happen in the generalized Kane-Mele-Hubbard model~\cite{Hung2013,Hung2014,Meng2014,He2015b} (GKMH), cluster Kane-Mele-Hubbard model~\cite{Wu2012,Grandi2015,He2015b} (CKMH), the bilayer Kane-Mele-Hubbard model (BKMH) with inter-layer antiferromagnetic spin-spin interaction $J$ (BKMH-$J$)~\cite{He2015a}, and inter-layer interaction $V$ (BKMH-$V$). The TPTs in the first two models have noninteracting correspondences, and by means of the periodization scheme developed in paper I~\cite{He2015b}, we obtain the ideally quantized topological invariants (spin Chern number) to characterize them. The TPTs in the second two models do not have noninteracting analogues, we found an unexpected breakdown of the Green's function formalism in constructing the topological invariants, in which it is clearly seen that after the interaction-driven TPTs, the trivial insulators (product states of $SO(4)$ inter-layer $J$-singlet or $V$-singlet) are still incorrectly associated with ideally quantized, nonzero spin Chern numbers if they were constructed from single-particle Green's function. These artificial results show the limitation of the single-particle Green's function formalism in monitoring interaction-driven TPTs and highlight calling for more complete understanding and versatile technique in studying interacting TIs.

The rest of the paper is organized as follows. In Sec.~\ref{sec:ApplyQMC}, the QMC results of the interaction-driven TPTs are presented, containing the ones in GKMH and CKMH models which have noninteracting correspondences and the Green's function formalism succeeds, and the others in BKMH-$J$ and BKMH-$V$ models which do not have noninteracting analogues and the Green's function formalism fails. In Sec.~\ref{sec:TopoCondition}, we carry out detailed analysis for the reason of the breakdown of spin Chern number constructed from Green's function formalism in characterizing certain interaction-driven TPTs. Finally, summary is given in Sec.~\ref{sec:Summary}.

\section{Numerical Results}
\label{sec:ApplyQMC}

In this section, we present the QMC simulation results for the four interaction-driven TPTs with neither explicit nor spontaneous symmetry breaking, with special focus on the numerical data of topological invariants calculated from the zero-frequency single-particle Green's function. As for the Green's function formalism in constructing the topological invariants, the periodization scheme to overcome the finite size effect in QMC simulation, and the basic introduction of the projector QMC technique itself, the readers are referred to paper I~\cite{He2015b}.

\subsection{Interaction-driven TPTs in GKMH model}
\label{sec:GKMH}

\begin{figure}[tp!]
\centering
\includegraphics[width=\columnwidth]{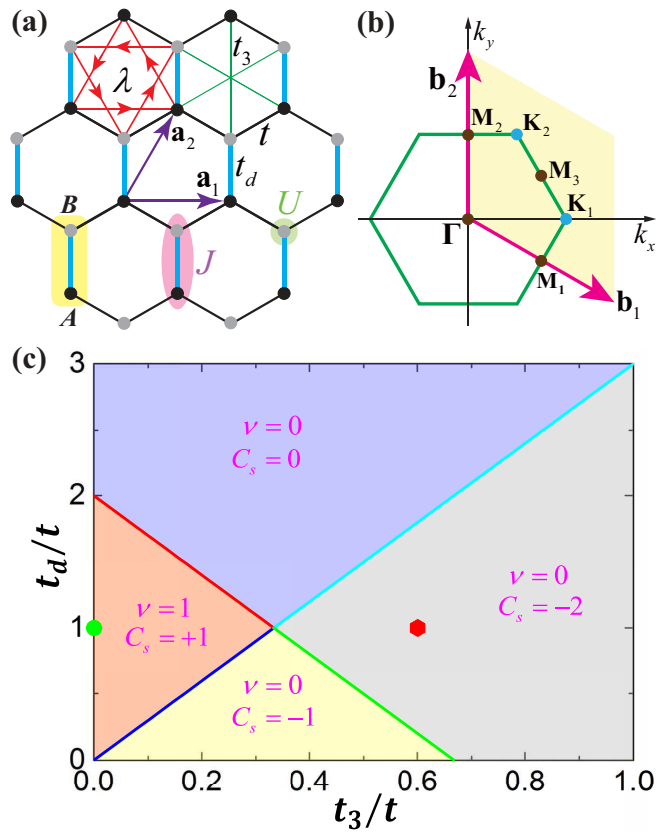}
\caption{\label{fig:KMHLatt}(Color online) (a) Illustration of honeycomb lattice and all the terms in GKMH model as Eq.~\ref{eq:GKMHModel}. The unit cell is presented as the yellow shaded rectangle, consisting of A and B sublattices denoted by black and gray circles. The lattice is spanned by primitive vectors $\mathbf{a}_1=(\sqrt{3},0)a$, $\mathbf{a}_2=(1/2,\sqrt{3}/2)a$ with $a$ the lattice constant. The black and blue lines denotes nearest-neighbor hopping between ($t$) and inside ($t_{d}$) unit cells, while the SOC term ($\lambda$) and third-neighbor hopping ($t_{3}$) are represented by red and green lines. The arrows in red lines shows $\nu_{ij}=+1$ for spin-up part. The on-site $U$ interaction and the AFM $J$ interaction existing only inside unit cell are presented by green circles and magenta ellipse. (b) The BZ of GKMH model. $\mathbf{K}_1,\mathbf{K}_2$ are Dirac points, while $\mathbf{\Gamma}$, $\mathbf{M}_1$, $\mathbf{M}_2$, $\mathbf{M}_3$ are the four TRI points. (c) The noninteracting $(t_d/t)-(t_3/t)$ phase diagram for GKMH model, and the $Z_2$ invariant $(-1)^\nu$ and spin Chern number $C_s$ for all the phases. The green dot ($t_d=t,t_3=0$) and red hexagon ($t_d=t,t_3=0.6t$) along with $\lambda=0.2t$ and $J=1$ are the chosen parameters, based on which we identify the $U$-driven TPTs in Fig.~\ref{fig:GKMH1Z2ChernU} and Fig.~\ref{fig:GKMH2Z2ChernU}, respectively.}
\end{figure}

As introduced in paper I~\cite{He2015b}, the generalized Kane-Mele-Hubbard (GKMH) model~\cite{Hung2013,Hung2014,Meng2014} is given by
\begin{eqnarray}
\hat{H} &=& -\sum_{\langle i,j\rangle\sigma}t_{ij}(c^{\dagger}_{i\sigma}c_{j\sigma} + h.c) -t_3\sum_{\langle\!\langle\!\langle i,j\rangle\!\rangle\!\rangle\sigma}(c^{\dagger}_{i\sigma}c_{j\sigma} + h.c.) \nonumber\\
&& +i\lambda\sum_{\langle\!\langle i,j \rangle\!\rangle \alpha\beta}v_{ij}(c^{\dagger}_{i\alpha}\sigma^{z}_{\alpha\beta}c_{j\beta}
-c^{\dagger}_{j\beta}\sigma^{z}_{\beta\alpha}c_{i\alpha}) \nonumber\\
&& +\frac{U}{2}\sum_{i}(n_{i\uparrow}+n_{i\downarrow}-1)^2    \nonumber \\
&& + \frac{J}{8}\sum_{\langle i,j \rangle}\big[(D_{i,j}-D^{\dagger}_{i,j})^2 -(D_{i,j}+D^{\dagger}_{i,j})^2\big],
\label{eq:GKMHModel}
\end{eqnarray}
where $D_{i,j}=\sum_{\sigma}c^{\dagger}_{i\sigma}c_{j\sigma}$. For the nearest-neighbor (NN) hopping, we have $t_{ij}=t_d$ for the NN bonds inside unit cells and $t_{ij}=t$ for the others, as shown in Fig.~\ref{fig:KMHLatt} (a). The $t_{3}$ term is the third-nearest-neighbor hopping. The $\lambda$ term represents spin-orbit coupling (SOC) connecting the next-nearest-neighbor sites with a complex (time-reversal symmetric) hopping with amplitude $\lambda$. The factor $\nu_{ij}=-\nu_{ji}=\pm1$ depends on the orientation of the two nearest-neighbor bonds that an electron moves in going from site $i$ to $j$. The $U$ term is the on-site Coulomb repulsion, while the $J$ term, which only exists for the NN bonds inside unit cells, is a faithful approximation~\cite{He2015a} of the antiferromagnetic (AFM) Heisenberg interaction $J\sum_{\langle ij \rangle}\mathbf{S}_{i}\cdot\mathbf{S}_{j}$. Throughout this paper, we set $t$ as the energy unit. The honeycomb lattice and its BZ are shown in Fig.~\ref{fig:KMHLatt} (a) and (b).

\begin{figure}[tp!]
\centering
\includegraphics[width=\columnwidth]{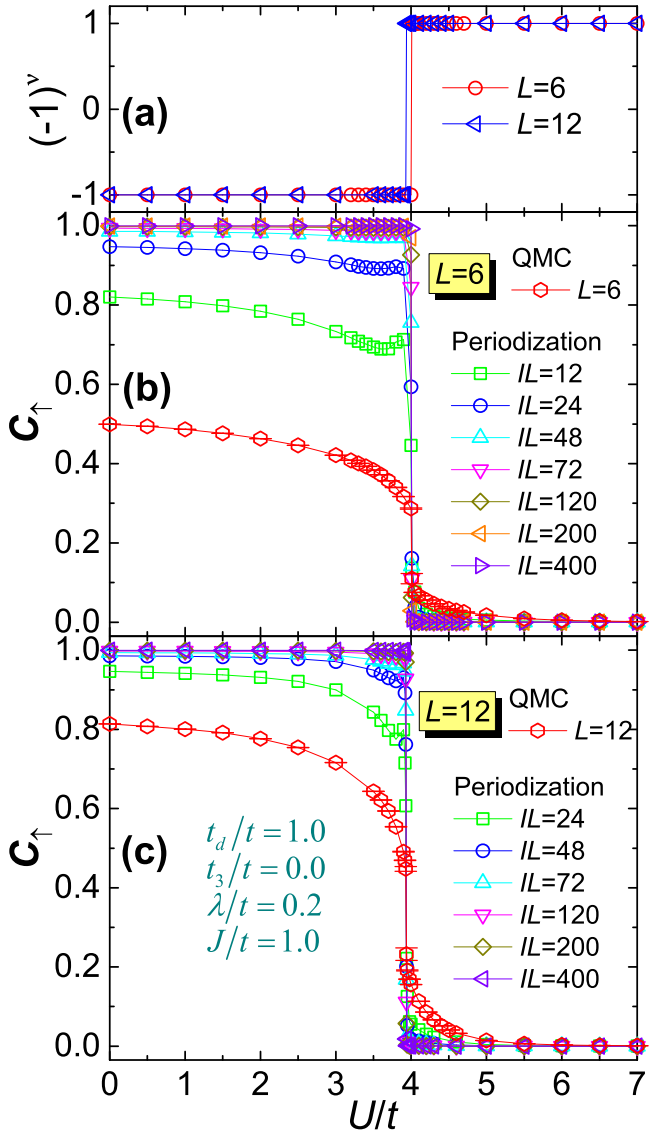}
\caption{\label{fig:GKMH1Z2ChernU}(Color online) (a) $Z_2$ invariant $(-1)^\nu$ and (b), (c) Chern number $C_{\uparrow}$ for the $U$-driven topological phase transition in GKMH model with $t_d=t,\lambda=0.2t,t_3=0$ (green dot in Fig.~\ref{fig:KMHLatt} (c)) and $J=t$ from QMC results for $L=6$ and $L=12$ and interpolation to large $IL$. We can observe that $Z_2$ invariant $(-1)^\nu$ are integer-quantized. The Chern number $C_{\uparrow}$ obtained from finite size systems $L=6,12$ acquire a jump with finite value at the transition point, but the jump is not ideally quantized due to finite size effect. After periodization, based on the $L=6,12$ QMC data, the ideally quantized Chern number is achieved.}
\end{figure}

Due to $U(1)_{\text{charge}}\times U(1)_{\text{spin}}\rtimes Z_2^T$ symmetry, the GKMH model without interaction belongs to $\mathbb{Z}$ classification. Fig.~\ref{fig:KMHLatt} (c) shows the $(t_d/t)-(t_3/t)$ phase diagram for $\lambda/t>0$, determined from both $Z_2$ invariant $(-1)^\nu$ and spin Chern number $C_s$. Here the phase boundaries is independent of $\lambda/t$ parameter as long as $\lambda/t>0$. As we see, there are four different phases with different spin Chern number $C_s$, in which three of them with spin Chern numbers $C_s\neq0$ are topologically nontrivial and the last one with $C_s=0$ is trivial.

Here we concentrate on two $U$-driven TPTs, which are described by spin Chern number variation in odd or even integer across the transitions. The anisotropy introduced by the $J$-term inside unit cell suppresses the $xy$-AFM long range order, which otherwise arises in the large $U$ limit with $J=0$~\cite{DZheng2011,Hohenadler2012,Wu2015}, but favors a topologically trivial dimerized insulator phase without breaking time-reversal symmetry and spin $U(1)$ symmetry. The $U$-driven TPT with spin Chern number variation $|\Delta C_s|=1$ can be realized by setting $t_d=t,\lambda=0.2t,t_3=0$ (green dot in Fig.~\ref{fig:KMHLatt} (c)), $J=t$ and increasing $U$. The other $U$-driven TPT with $|\Delta C_s|=2$ appears by choosing $t_d=t,\lambda=0.2t,t_3=0.6t$ (red hexagon in Fig.~\ref{fig:KMHLatt} (c)), $J=t$.
For both phase transitions, the possible intervening $xy$-AFM long-range order is excluded by extrapolating the corresponding magnetic structure factor to thermodynamic limit. Thus, there is no spontaneous symmetry breaking during the transitions. We also need to emphasize that such interaction-driven TPTs can exist for a fairly large range of parameters, as long as the intervening $xy$-AFM order is absent, and we only demonstrate the above two representative cases.

For the TPT with $|\Delta C_s|=1$ in GKMH model, the results of both $Z_2$ invariant and spin Chern number $C_s$ are shown in Fig.~\ref{fig:GKMH1Z2ChernU}. The integer-quantized $Z_2$ invariant determines the interaction-driven TPT with $U_c \approx 4.005t$ for $L=6$ system and $U_c\approx 3.935t$ for $L=12$ system, as shown in Fig.~\ref{fig:GKMH1Z2ChernU} (a), which indicates a very small finite-size effect in the topological phase transition point. We find that across the transition both the parity changing and the single-particle gap closing only happen at the $\mathbf{M}_2$ point in the BZ, which is due to the anisotropy introduced by the $J$-term. However, the Chern number $C_{\uparrow}$ calculated from finite size systems $L=6,12$ shown in Fig.~\ref{fig:GKMH1Z2ChernU} (b) and (c) suffers from severe finite-size effect, meanwhile the value and jump in $C_s$ are far from ideally quantization, even though its finite-value jumps happen at the transition points. We then apply the periodization scheme~\cite{He2015b} to obtain its integer-quantized values. The results are also shown in Fig.~\ref{fig:GKMH1Z2ChernU} (b) and (c). We have carried out interpolation for QMC results on $L=6$ (Fig.~\ref{fig:GKMH1Z2ChernU} (b)) and $L=12$ (Fig.~\ref{fig:GKMH1Z2ChernU} (c)) systems and the interpolation lattice size $IL$ can be as large as $IL=400$. The convergence of $C_{\uparrow}$ with $IL$ to its expected integer value can be clearly observed.

Combining the $Z_2$ invariant $(-1)^\nu$ and spin Chern number $C_s=C_{\uparrow}$ in Fig.~\ref{fig:GKMH1Z2ChernU}, the $U$-driven, QSH insulator to dimer insulator transition is clearly established. In the dimer insulator phase, spin singlets are formed on the bonds inside unit cell due to presence of $J$ and $U$ interactions. Under the dimer limit, the system is actually a direct product state, which is topologically trivial $(C_s=0)$ since all the electronic degrees of freedom are frozen and there is no edge state even if open boundary was created. This trivial dimer insulator is adiabatically connected -- without going through phase transition and symmetry breaking -- to the $\nu=0,C_s=0$ noninteracting trivial insulator in Fig.~\ref{fig:KMHLatt} (c), while the interacting quantum spin Hall (QSH) insulator at $J=1$ and $U<4t$ with $C_s=1$ is adiabatically connected to the $\nu=1,C_s=+1$ phase in Fig.~\ref{fig:KMHLatt} (c). This means that the above $U$-driven TPT with $|\Delta C_s|=1$ has a noninteracting correspondence which is the transition at the red line in Fig.~\ref{fig:KMHLatt} (c) and the effect of interactions is only to renormalize the hopping parameters.

\begin{figure}[tp!]
\centering
\includegraphics[width=\columnwidth]{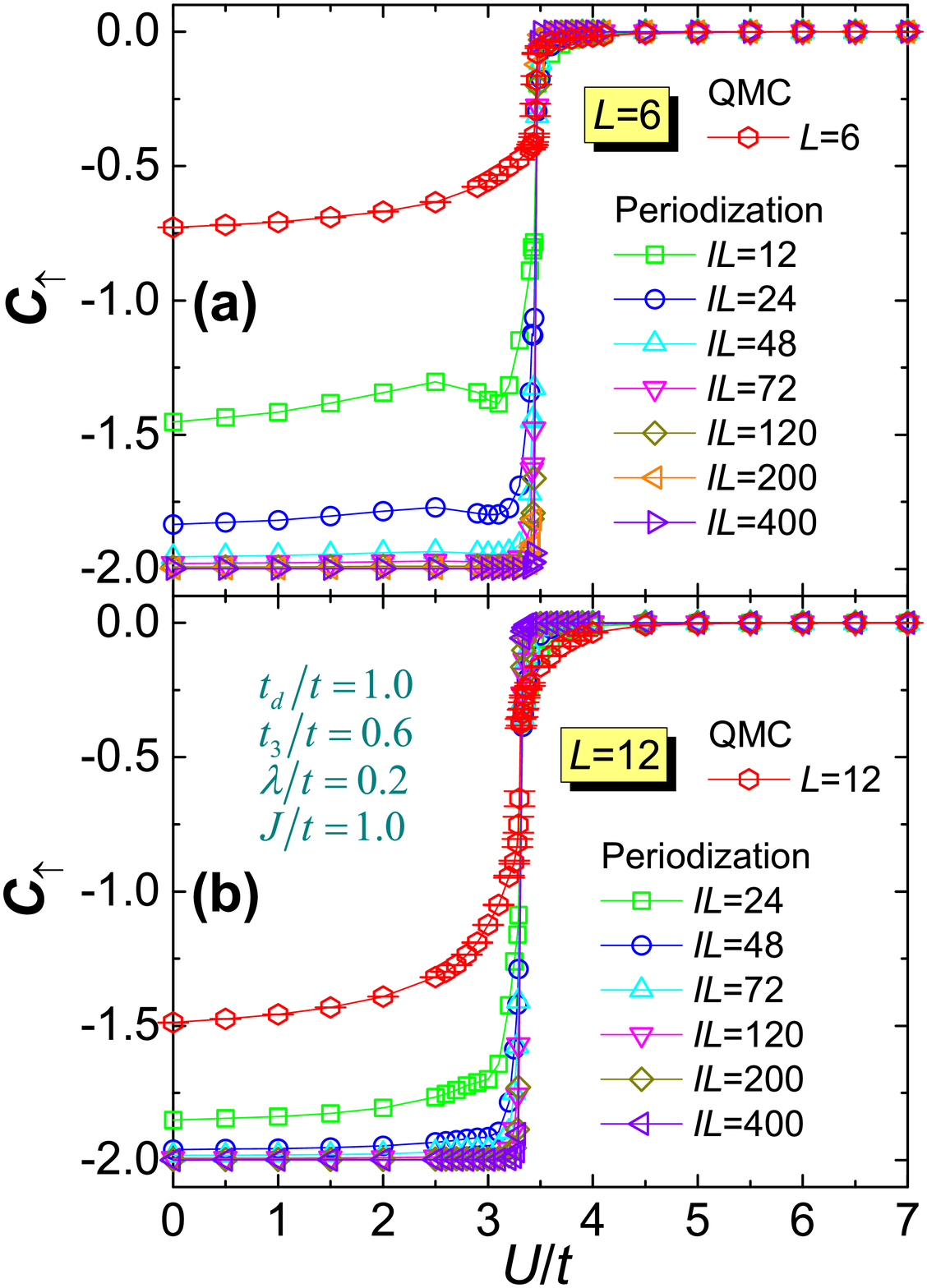}
\caption{\label{fig:GKMH2Z2ChernU}(Color online) (a), (b) Chern number $C_{\uparrow}$ for the $U$-driven topological phase transition in GKMH model with $t_d=t,\lambda=0.2t,t_3=0.6t$ (red hexagon in Fig.~\ref{fig:KMHLatt} (c)) and $J=t$ from finite-size calculations by QMC method for $L=6$ and $L=12$ systems and periodization with large $IL$. For finite size system, the Chern number $C_{\uparrow}$ accquires a jump with finite value which is not quantized at the transition point, after the periodization, the convergence to the ideally quantized Chern number is clearly seen and it sharply defines the interaction-driven topological phase transition.}
\end{figure}

For the TPT with $|\Delta C_s|=2$ in GKMH model, the results of spin Chern number $C_s$ are shown in Fig.~\ref{fig:GKMH2Z2ChernU}, for both finite size QMC and periodization results. Without interaction, the system is a TI with $C_s=-2$, for which the $Z_2$ invariant is $\nu=0$ as $(-1)^\nu=+1$. With $J=t$ and increasing $U$, a $U$-driven TPT from $C_s=-2$ TI to $C_s=0$ dimer insulator is expected. Across this TPT, we observe that the single-particle gap closing and the parity change happen at both $\mathbf{M}_1$ and $\mathbf{M}_3$ points, although the total $Z_2$ parity does not have a variation. Combining the jumps in both parity and Chern number $C_{\uparrow}$ with the gap closing behavior, we determine the transition points for this TPT as $U_c\approx 3.46(5)t$ for $L=6$ and $U_c\approx 3.30(5)t$ for $L=12$, which gives a $\Delta U_c\approx 0.16t$ shift of the phase boundary due to finite size effect in QMC.

To reach ideally quantized spin Chern number for this $U$-driven TPT, the periodization scheme is then applied. The calculation results of Chern number $C_{\uparrow}$ after the periodization from $L=6,12$ systems are also shown in Fig.~\ref{fig:GKMH2Z2ChernU} (a) and (b). Again, we achieve the ideally quantized Chern number $C_{\uparrow}$ and a sharp transition from $C_s=-2$ to $C_s=0$ when the interpolation lattice size $IL$ is large enough. There are some nonmonotonic behaviors in the results for small $IL$. These behaviors are caused by the $\tau$ cutoff that we applied in the periodization~\cite{He2015b}, but they disappear when $IL$ is large enough.

Similar to the $|\Delta C_s=1|$ case, this $U$-driven TI-dimer TPT with $|\Delta C_s|=2$ also has noninteracting correspondence. The $C_s=-2$ TI phase at $J=t$ and $U<3.5$ is adiabatically connected to the $\nu=0,C_s=-2$ TI phase in Fig.~\ref{fig:KMHLatt} (c), while the $\nu=0,C_s=0$ dimer-singlet insulator at $J=t$ and $U>3.5$ is adiabatically connected to the $\nu=0,C_s=0$ phase in Fig.~\ref{fig:KMHLatt} (c) as well. So the $U$-driven TPT in Fig.~\ref{fig:GKMH2Z2ChernU} is exactly the same as the transitions on the solid cyan line in Fig.~\ref{fig:KMHLatt} (c). Considering both TPTs with $|\Delta C_s|=1$ and $|\Delta C_s|=2$, it's interesting that the spin Chern number constructed from single-particle Green's function can detect interaction-driven TPTs, even though interaction $U$ for both transitions is large. Clearly, this is due to the fact that these TPTs have noninteracting correspondences.

\subsection{Interaction-driven TPTs in CKMH model}
\label{sec:CKMH}

As introduced in paper I~\cite{He2015b}, the CKMH model~\cite{Grandi2015,Wu2012} has 6 lattice sites per unit cell, and the model Hamiltonian is given by
\begin{eqnarray}
\hat{H} =&& -\sum_{\langle ij\rangle\sigma}t_{ij}(c^{\dagger}_{i\sigma}c_{j\sigma} + c^{\dagger}_{j\sigma}c_{i\sigma})  \nonumber\\
&& +i\lambda_{I}\sum_{\langle\!\langle ij \rangle\!\rangle \alpha\beta}v_{ij}(c^{\dagger}_{i\alpha}\sigma^{z}_{\alpha\beta}c_{j\beta}
-c^{\dagger}_{j\beta}\sigma^{z}_{\beta\alpha}c_{i\alpha}) \nonumber\\
&& +i\lambda_{O}\sum_{\langle\!\langle ij \rangle\!\rangle \alpha\beta}v_{ij}(c^{\dagger}_{i\alpha}\sigma^{z}_{\alpha\beta}c_{j\beta}
-c^{\dagger}_{j\beta}\sigma^{z}_{\beta\alpha}c_{i\alpha}) \nonumber\\
&& +\frac{U}{2}\sum_{i}(n_{i\uparrow}+n_{i\downarrow}-1)^2\;.
\label{eq:CKMH}
\end{eqnarray}
\begin{figure}[tp!]
\centering
\includegraphics[width=0.96\columnwidth]{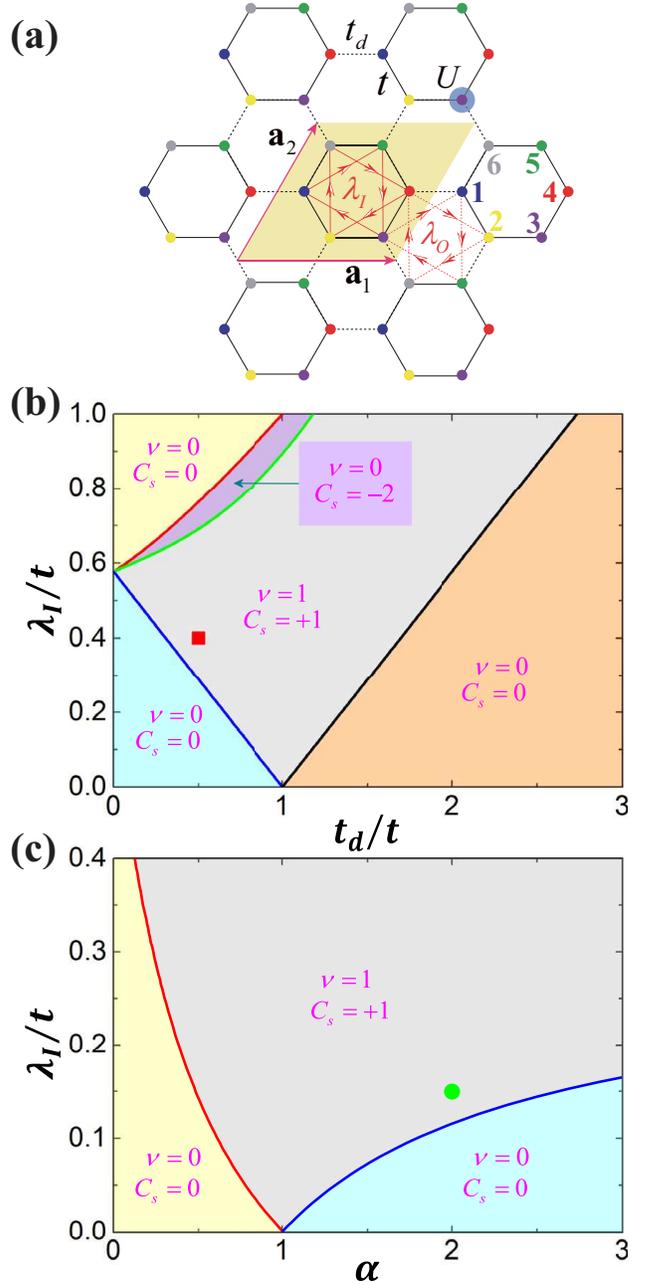}
\caption{\label{fig:ClusterKMH}(Color online) (a) The CKMH model with 6-six site unit cell. The yellow shaded region shows the unit cell with primitive lattice vectors $\mathbf{a}_1=(\sqrt{3},0)a$, $\mathbf{a}_2=(1/2,\sqrt{3}/2)a$ while the distance between nearest-neighbor lattice sites is $a/\sqrt{3}$. The black solid and dotted lines indicate the nearest-neighbor hopping terms inside and between different unit cells. The red solid and dotted lines represent the SOC terms inside and between different unit cells. The sign choice for SOC hopping is the same as that in Fig.~\ref{fig:KMHLatt} (a). The on-site Coulomb repulsion is shown by the blue shaded circle. (b) The noninteracting $(t_d/t)-(\lambda_I/t)$ phase diagram for CKMH model with $\lambda_O=0$. (c) The noninteracting $\alpha-(\lambda_I/t)$ phase diagram for CKMH model. The red square ($t_d/t=0.5,\lambda_I/t=0.4$) and green dot ($\alpha=0.5,\lambda_I/t=0.15$) are the chosen parameters based on which the $U$-driven TPTs are studied in Fig.~\ref{fig:CKMH1Z2ChernU} and Fig.~\ref{fig:CKMH2Z2ChernU}, respectively.}
\end{figure}
For the nearest-neighbor (NN) hopping, we have $t_{ij}=t$ for the NN bonds inside unit cells and $t_{ij}=t_d$ for those connecting the six-site unit cells, as shown in Fig.~\ref{fig:ClusterKMH} (a). $\lambda_I$ and $\lambda_O$ are the SOC terms inside and between unit cells, respectively. $U$ is the on-site Coulomb repulsion. Similar to the GKMH model, the CKMH model preserves $U(1)_{\text{change}}\times U(1)_{\text{spin}}\rtimes Z_2^T$ symmetry, which also results in $\mathbb{Z}$ classification. Besides, both spatial inversion symmetry and particle-hole symmetry are also conserved in CKMH model. The hexagonal BZ of the CKMH model differs with that of the GKMH model in Fig.~\ref{fig:KMHLatt} (b) only up to a rescaling of reciprocal lattice vectors.

Similar to the CKMH model discussed in paper I~\cite{He2015b}, we first set $\lambda_O=0$ and only keep the SOC term $\lambda_I$ finite. In this case, the $(t_d/t)-(\lambda_I/t)$ phase diagram of noninteracting CKMH model is shown in Fig.~\ref{fig:ClusterKMH} (b). We can observe that two TIs exist in middle region of the phase diagram, with different spin Chern numbers $C_s=+1$ at small $\lambda_I$ and $C_s=-2$ at larger $\lambda_I$. Then, we keep all three hopping parameters $t_d$, $\lambda_I$, and $\lambda_O$ finite and introduce a ratio of $\alpha=t_d/t=\lambda_O/\lambda_I$. The noninteracting $\alpha-(\lambda_I/t)$ phase diagram is presented in Fig.~\ref{fig:ClusterKMH} (c).

In CKMH model, a large on-site $U$ interaction can drive the system into a topologically trivial insulator without spontaneous symmetry breaking, i.e., without AFM long range order, provided that the value of $t_d/t$ is far away from $1$. As will become clear below, these trivial insulators can be either a plaquette valance bond solid (pVBS) or columnar valance bond solid (cVBS)~\cite{Lang2013}. These VBSs are insulators built from spin singlets, with either two electrons or six electrons. In the following, we first study the $U$-driven TPT with $t_d=0.5t, \lambda_I=0.4t$ (red square in Fig.~\ref{fig:ClusterKMH} (b)), and then the $U$-driven TPT with $\alpha=2,\lambda_I=0.15t$ (green dot in Fig.~\ref{fig:ClusterKMH} (c)).

\begin{figure}[tp!]
\centering
\includegraphics[width=\columnwidth]{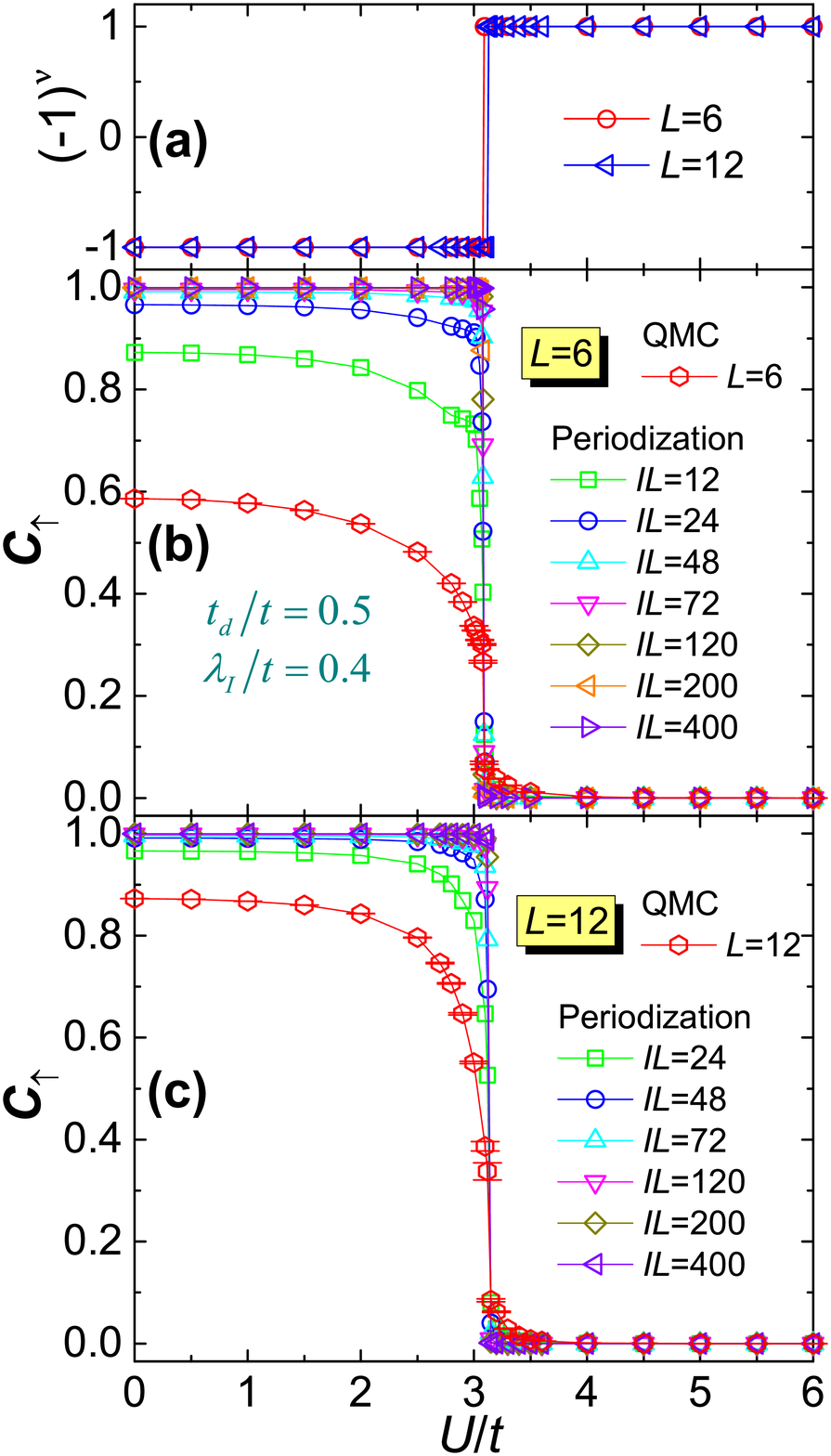}
\caption{\label{fig:CKMH1Z2ChernU}(Color online) (a) $Z_2$ invariant $(-1)^\nu$ and (b), (c) Chern number $C_{\uparrow}$ for the $U$-driven topological phase transition in CKMH model with $t_d=0.5t, \lambda_I=0.4t$ (red square in Fig.~\ref{fig:ClusterKMH} (b)) from QMC calculation for $L=6,12$ system and periodization with large $IL$. Across the TPT point, the quantized $Z_2$ invariant experiences an integer variation and Chern number $C_{\uparrow}$ acquires a finite-value jump. After the periodization with the QMC data in $L=6,12$ with large $IL$, the Chern number $C_{\uparrow}$ reaches its ideally quantized value.}
\end{figure}

We first study the $U$-driven TPT marked by the red square $t_d=0.5t, \lambda_I=0.4t$ in Fig.~\ref{fig:ClusterKMH} (b). Results of both $Z_2$ invariant $(-1)^\nu$ and Chern number $C_{\uparrow}$ from the QMC data with $L=6,12$ and periodization with large $IL$ are presented in Fig.~\ref{fig:CKMH1Z2ChernU}. Both the integer variation of $Z_2$ invariant (Fig.~\ref{fig:CKMH1Z2ChernU} (a)) and finite value jump of Chern number $C_{\uparrow}$ (Fig.~\ref{fig:CKMH1Z2ChernU} (b) and (c)) suggest the TPT at $U_c\approx 3.085t$ for $L=6$ system and $U_c\approx 3.125t$ for $L=12$ system. Across the transition, we find that both single-particle gap closing and parity change happen at $\mathbf{\Gamma}$ point. To obtain the ideally quantized Chern number $C_{\uparrow}$, the periodization scheme is applied, as shown in Fig.~\ref{fig:CKMH1Z2ChernU} (b) and (c). The quantized $C_{\uparrow}$ results in Fig.~\ref{fig:CKMH1Z2ChernU} (b) and (c) further confirm the $U$-driven TPT with spin Chern number variation $|\Delta C_s|=1$.

The topologically trivial phase after the TPT is a pVBS, in which six electrons form a total spin singlet inside a unit cell indicated by the yellow shaded region in Fig.~\ref{fig:ClusterKMH} (a). This phase is easy to understand. With large $U$, the effective model of the system becomes $J-J^\prime$ Heisenberg model in which $J,J^\prime$ are inside and between different unit cells with $J\propto t^2/U$ and $J^\prime\propto t_d^2/U$, respectively. For the parameter $t_d=0.5t$, we arrive at approximately $J\approx 4J^\prime$ (neglecting the contribution of SOC hopping term), such $J-J^\prime$ Heisenberg model acquires a pVBS ground state on honeycomb lattice~\cite{Lang2013}. Thus this TPT is a $U$-driven QSH-to-pVBS transition. What's more, the $(-1)^\nu=-1,C_s=+1$ QSH insulator at $U<3$ in CKMH model is adiabatically connected to the $\nu=1,C_s=+1$ grey-colored QSH region in the noninteracting phase diagram Fig.~\ref{fig:ClusterKMH} (b). In contrast, the $(-1)^\nu=+1,C_s=0$ phase for $U>3$ is adiabatically connected to the $\nu=0,C_s=0$ phase in the cyan region in Fig.~\ref{fig:ClusterKMH} (b). Hence the TPT shown in Fig.~\ref{fig:CKMH1Z2ChernU} is exactly the same as the transition along the blue solid line in the noninteracting phase diagram Fig.~\ref{fig:ClusterKMH} (b).

\begin{figure}[htp!]
\centering
\includegraphics[width=\columnwidth]{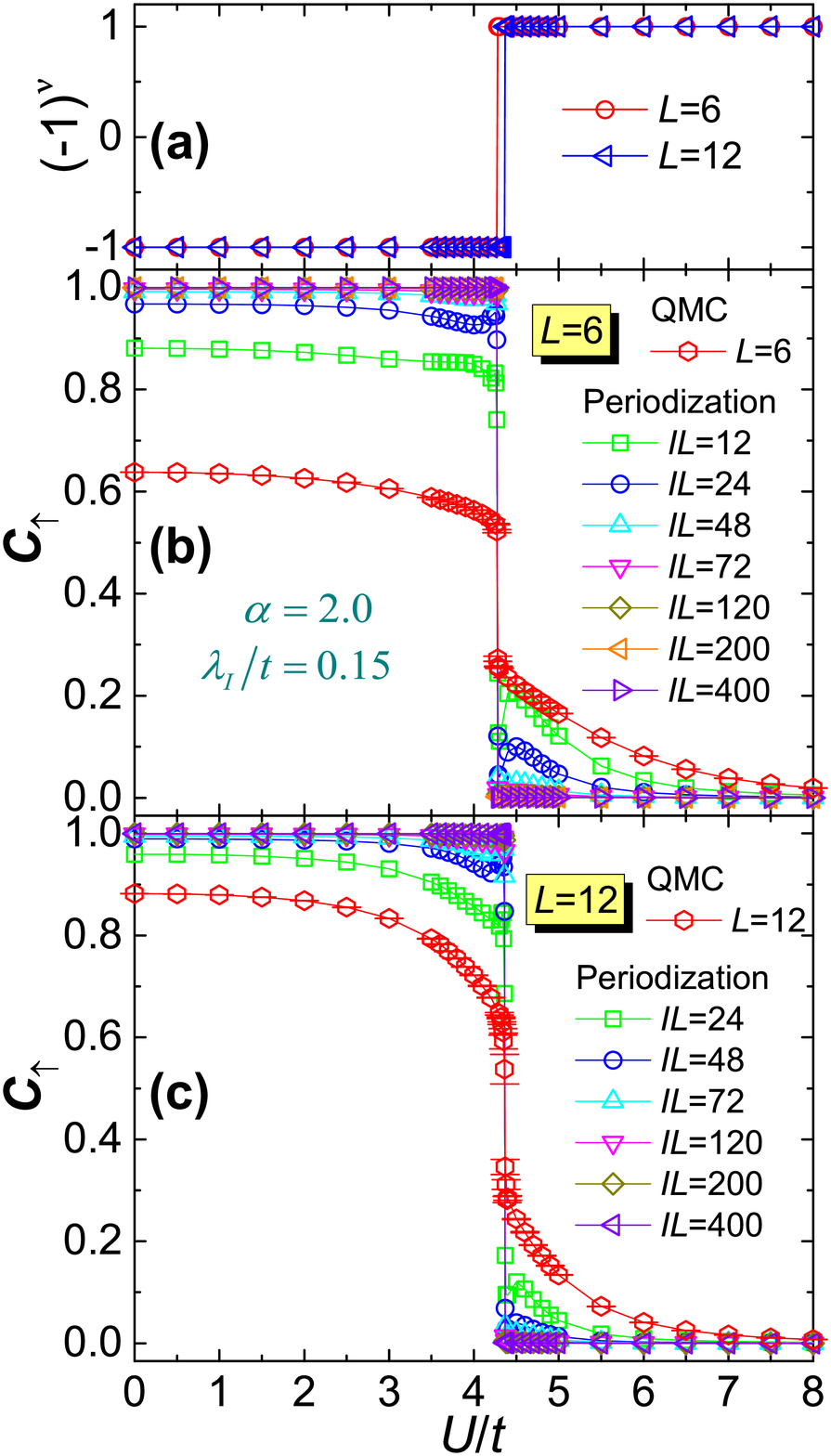}
\caption{\label{fig:CKMH2Z2ChernU}(Color online) (a) $Z_2$ invariant $(-1)^\nu$ and (b), (c) Chern number $C_{\uparrow}$ for the $U$-driven topological phase transition in CKMH model with $\alpha=2, \lambda_I=0.15t$ (green dot in Fig.~\ref{fig:ClusterKMH} (c)) from QMC calculation for $L=6,12$ systems and periodization with large $IL$. Across the TPT, the quantized $Z_2$ invariant experiences an integer variation and Chern number $C_{\uparrow}$ accquires a finite-value jump. After the periodization, the Chern number $C_{\uparrow}$ reaches ideally quantized value and demonstrates the interaction-driven TPT.}
\end{figure}

We then study the $U$-driven TPT in CKMH model starting from the green dot $\alpha=2, \lambda_I=0.15t$ in Fig.~\ref{fig:ClusterKMH} (c). Results of both $Z_2$ invariant $(-1)^\nu$ and Chern number $C_{\uparrow}$ calculated from the QMC data in $L=6,12$ systems and periodization with large $IL$ are respectively shown in Fig.~\ref{fig:CKMH2Z2ChernU} (a), (b) and (c). Both these two topological invariants experience jumps at $U_c\approx 4.275t$ for $L=6$ system and $U_c\approx 4.365t$ for $L=12$ system, suggesting the transition point $U_c\approx 4.37t$. Across the transition, we also observe that the single-particle gap closing and parity change happen at $\mathbf{\Gamma}$ point. Likewise, the periodization scheme, applied upon the QMC data in $L=6$ and $L=12$ systems, gives the ideally quantized C across this TPT, as shown in Fig.~\ref{fig:CKMH2Z2ChernU} (b) and (c).

With $\alpha=2, \lambda_I=0.15t$, at large $U$ limit, in the effective $J-J^\prime$ model, we now have $J^\prime > J$, so the ground state of CKMH model at $U>4.4$ is a cVBS state~\cite{Lang2013}, in which the spin singlets form on the $t_{d}$ bonds in Fig.~\ref{fig:ClusterKMH} (a) connecting different unit cells. Similar to the CKMH model with $t_d=0.5t, \lambda_I=0.4t$ discussed above, the $(-1)^\nu=-1,C_s=+1$ phase at $U<4.4$ is adiabatically connected to the grey-colored $\nu=1,C_s=+1$ QSH region in noninteracting phase diagram Fig.~\ref{fig:ClusterKMH} (c), while the cVBS phase at $U>4.4$ is adiabatically connected to the $\nu=0,C_s=0$ phase with light cyan color in Fig.~\ref{fig:ClusterKMH} (c). So this $U$-driven QSH-to-cVBS TPT is the same as the transition on the blue solid line in Fig.~\ref{fig:ClusterKMH} (c).

\subsection{Interaction-driven TPT in BKMH-$J$ model}
\label{sec:BKMHJ}

The next two interaction-driven TPTs are very different from those discussed above, in that, from here we will observe breakdown of the topological invariants constructed from the Green's function formalism, and find new type of interaction-driven TPT where the fermions are gapped throughout the TPT but there are emergent collective bosonic modes that become critical at the transition. The topological trivial phases after the TPT are featureless Mott insulators, which do not have noninteracting correspondences.

\begin{figure}[tp!]
\centering
\includegraphics[width=\columnwidth]{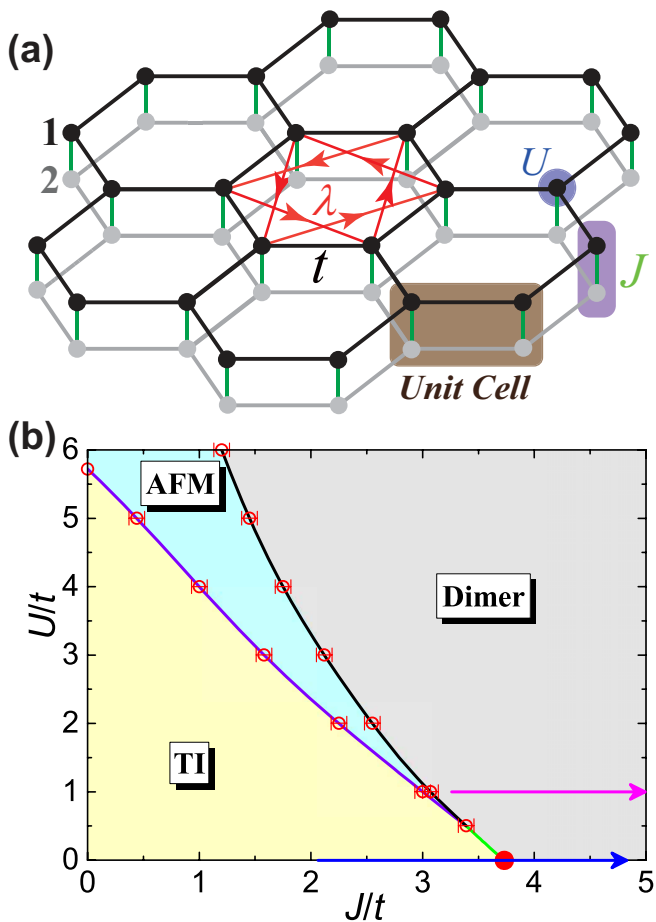}
\caption{\label{fig:BKMHLatticeGeometry} (Color online) (a) Illustration of AA-stacked honeycomb lattice and bilayer KMH model with inter-layer AFM interaction $J$. The four-site unit cell is presented as the shaded rectangle. The gray and black lines indicates the nearest-neighbor hopping $t$ on layer 1 and 2, respectively. The spin-orbital coupling term $\lambda$, for one spin flavor, is shown by the red lines and arrows with $\nu_{ij}=+1$. The on-site Coulomb repulsion $U$ and inter-layer AFM coupling $J$ are represented by the shaded circle and rectangle. (b) The $U$-$J$ phase diagram for BKMH-$J$ model for $\lambda=0.2t$. The blue and magenta lines with arrow demonstrate the paths where the topological invariants constructed from Green's function are calculated in Fig.~\ref{fig:BKMHChernU0Period} and Fig.~\ref{fig:BKMHChernU1Period}, respectively.}
\end{figure}

Let's begin with the bilayer Kane-Mele-Hubbard model with inter-layer AFM interaction $J$~\cite{Slagle2015,He2015a,You2015}. The Hamiltonian of BKMH-$J$ model is given as
\begin{eqnarray}
H =&& -t\sum_{\xi\langle i,j
\rangle,\alpha}( c^{\dagger}_{\xi i\alpha}c_{\xi j\alpha} + c^{\dagger}_{\xi j\alpha}c_{\xi i\alpha} )  \nonumber \\
&&+ i\lambda\sum_{\xi\langle\!\langle i,j \rangle\!\rangle, \alpha\beta}
v_{ij}(c^{\dagger}_{\xi i\alpha}\sigma^{z}_{\alpha\beta}c_{\xi j\beta} - c^{\dagger}_{\xi j\beta}\sigma^{z}_{\beta\alpha}c_{\xi i\alpha}) \nonumber \\
  &&+ \frac{U}{2}\sum_{\xi i}(n_{\xi i}-1)^2 \nonumber \\
  &&+ \frac{J}{8}\sum_{i}\big[(D_{1i,2i}-D^{\dagger}_{1i,2i})^2 -(D_{1i,2i}+D^{\dagger}_{1i,2i})^2\big], \hspace{0.5cm}
 \label{eq:ModelHamiltonian-BKMH-J}
\end{eqnarray}
where $\xi=1,2$ is the layer index. Inside each layer, the nearest-neighbor hopping $t$ and SOC term $\lambda$ are the same as those in GKMH model. The interactions include the on-site Coulomb repulsion $U$ and inter-layer AFM exchange coupling $J$, which has the same expression as the $J$ interaction term in Eq.~\ref{eq:GKMHModel}, except that now the $J$-term is inter-layer one. The lattice geometry and all the terms in BKMH-$J$ Hamiltonian are presented in Fig.~\ref{fig:BKMHLatticeGeometry} (a). Its unit cell contains four lattice sites, and the corresponding BZ is the same as that in Fig.~\ref{fig:KMHLatt} (b).

The BKMH-$J$ model has $U(1)_{\text{spin}}\times[U(1)\times U(1)]_{\text{charge}}\rtimes Z_2^T$ symmetry. Here the two $U(1)$ charge symmetries correspond to the charge conservations in each layer. The detailed $U$-$J$ phase diagram of BKMH-$J$ model has already been carefully studied in Ref.~\onlinecite{He2015a} by the QMC simulations. With finite $\lambda$ and no interaction $U=J=0$, the system is a TI with spin Chern number $C_s=+2$ and $Z_2$ invariant $(-1)^\nu=+1$. With interaction, at large $U$ limit with small $J$, the system enters the $xy$-AFM phase~\cite{Hohenadler2012,Meng2014,He2015a}. With large $J$, inter-layer dimer-singlet insulator is naturally the ground state of BKMH-$J$ model. The $U$-$J$ phase diagram determined from the QMC simulations is presented in Fig.~\ref{fig:BKMHLatticeGeometry} (b) for $\lambda=0.2t$.

As demonstrated in Ref.~\onlinecite{He2015a,You2015}, the BKMH-$J$ model with $U=0$ has an $SO(4)\simeq SU(2)\times SU(2)$ symmetry. To see it more clearly, we define $f_{i\uparrow}=(c_{1i\uparrow},(-1)^ic_{2i\uparrow}^{\dagger})^T$ and $f_{i\downarrow}=((-1)^ic_{1i\downarrow},c_{2i\downarrow}^{\dagger})^T$, then the $SO(4)$ symmetry of BKMH-$J$ model at $U=0$ becomes explicit, since we can rewrite the model Hamiltonian with $\hat{P}_i=\frac{1}{2}(-1)^{i}\sum_{\sigma}f_{i\sigma}^{\dagger}i\tau^2(f_{i\sigma}^{\dagger})^{\text{T}}$ as
\begin{eqnarray}
\label{eq:BKMHQMCFfermion}
H = \sum_{i,j,\sigma}\chi_{\sigma}(f_{i\sigma}^{\dagger}t_{ij}f_{j\sigma} + h.c.)
              - \frac{J}{4}\sum_{i}(\hat{P}^{\dagger}_{i}\hat{P}_{i}+ \hat{P}_{i}\hat{P}^{\dagger}_{i}), \hspace{0.4cm}
\end{eqnarray}
where we have $\chi_{\sigma}=(-1)^{\sigma}$, and $t_{ij}=t$ for hoppings on the NN bonds and $t_{ij}=i\lambda$ for SOC on the next-nearest-neighbor (NNN) bonds. The Hamiltonian in Eq.~\ref{eq:BKMHQMCFfermion} is invariant under the transformation: $f_{i\sigma}\to U_{\sigma}f_{i\sigma}$ with $U_{\sigma}\in SU(2)$ for both $\sigma=\uparrow,\downarrow$, so the BKHM-$J$ model at $U=0$ indeed has the $SO(4)\simeq SU(2)\times SU(2)$ symmetry. This $SO(4)$ symmetry results in the degeneracy of inter-layer spin-singlet $s$-wave pairing order and the inter-layer $xy$-AFM order~\cite{He2015a,You2015}, such that both the inter-layer spin-singlet $s$-wave pairing gap and the inter-layer spin gap close at the $J$-driven TPT point denoted by the red point in Fig.~\ref{fig:BKMHLatticeGeometry} (b).

Here we focus on the $J$-driven TPT at $U=0$, as denoted by the blue line with arrow in Fig.~\ref{fig:BKMHLatticeGeometry} (b). From the QMC results in Ref.~\onlinecite{He2015a}, at the $J$-driven TPT point, both spin and charge gap close but the single-particle gap remains open, i.e., the fermionic degree of freedom is gapped out throughout the entire $J$-axis. This is in sharp contrast with the interaction-driven TPTs in both GKMH (Sec.~\ref{sec:GKMH}) and CKMH (Sec.~\ref{sec:CKMH}) models, as well as those one-body-parameter-driven TPTs discussed in paper I~\cite{He2015b}. This unique property means that this TPT has no free fermion correspondence.

\begin{figure}[tp]
\centering
\includegraphics[width=\columnwidth]{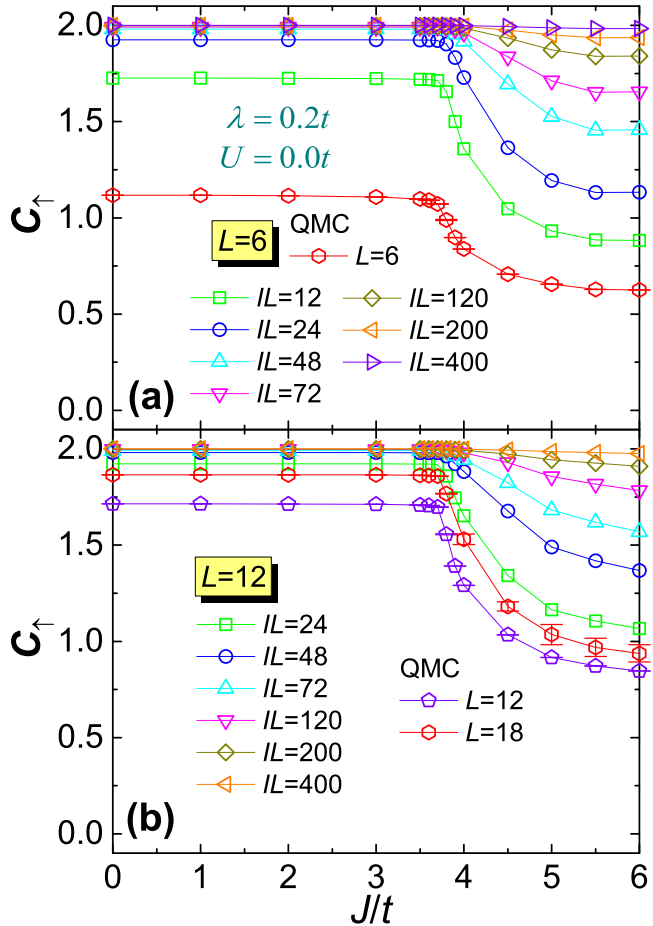}
\caption{\label{fig:BKMHChernU0Period}(Color online) Chern number $C_{\uparrow}$ for the $J$-driven TPT in BKMH-$J$ model with $\lambda=0.2t, U=0$ (the blue line with arrow in Fig.~\ref{fig:BKMHLatticeGeometry} (b)) from QMC with $L=6$, $12$ and $18$ and periodization scheme, using the QMC data in (a) $L=6$ and (b) $L=12$ systems. The ideally quantized Chern number $C_{\uparrow}$ in large $IL$ cases indicate that it has no variation across the $J$-driven TPT.}
\end{figure}

What is the situation if we still perform the spin Chern number calculation based on single-particle Green's function as in GKMH (Sec.~\ref{sec:GKMH}) and CKMH (Sec.~\ref{sec:CKMH}) models? At $U=0$ with increasing $J$, the QMC results of Chern number $C_{\uparrow}$ for $L=6$ and $L=12,18$ systems and the periodization results with large $IL$ are presented in Fig.~\ref{fig:BKMHChernU0Period} (a) and (b). As expected, the $L=6$, $12$ and $18$ results suffer severe finite-size effect and $C_{\uparrow}$ is much smaller than expected $C_s=+2$ even for very small $J$. After the periodization with large $IL$, the ideally quantized $C_{\uparrow}$ are obtained both in Fig.~\ref{fig:BKMHChernU0Period} (a) and (b), but, a very unexpected behavior appears: the $C_{\uparrow}$ constructed from single-particle Green's function do not change across this $J$-driven TPT! Correspondingly, we also observe that there is no parity change in any of the TRI points, and we find that the single-particle gap is also finite at the transition point, but it is the inter-layer singlet pairing gap and inter-layer spin gap that close at the critical point $J_c\simeq 3.73$, as shown in Ref.~\onlinecite{He2015a}. From Fig.~\ref{fig:BKMHChernU0Period}, it's really obscure to determine whether there is a $J$-driven $C_s=+2$ to $C_s=0$ transition. Yet, we know that when $J>J_c$, the system is inside the inter-layer dimer-singlet insulator phase without any edge states~\cite{He2015a}, as it is a direct product states of inter-layer $J$-singlets, under large-$J$ limit. This seemingly contradicting result actually points out that the spin Chern number constructed from Green's function fails in detecting the $J$-driven TPT in BKMH-$J$ model. Actually, the same results of spin Chern number have also been obtained in Ref.~\onlinecite{Slagle2015} with finite frequency single-particle Green's function. This failure of spin Chern number is closely related to the inter-layer dimer-singlet insulator phase, which has no free fermion correspondence, as well as the special nature of this TPT, i.e., it is the collective bosonic modes that become critical at the transition whereas the fermionic degree is always gapped~\cite{He2015a}.

\begin{figure}[t]
\centering
\includegraphics[width=\columnwidth]{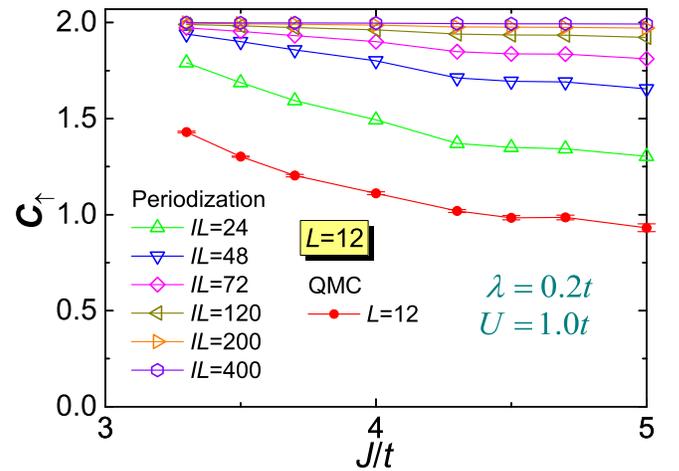}
\caption{\label{fig:BKMHChernU1Period}(Color online) Chern number $C_{\uparrow}$ for the finite-$U$ inside the $SO(4)$ $J$-singlet insulator phase in BKMH-$J$ model with $\lambda=0.2t, U=1$ (the magenta line with arrow in Fig.~\ref{fig:BKMHLatticeGeometry} (b)) from both finite-size QMC calculation and periodization method, using the QMC data of $L=12$ systems. Chern number $C_{\uparrow}$ converges to $C_{\uparrow}=+2$ with increasing $IL$.}
\end{figure}

To further illustrate the breakdown of the spin Chern number constructed from Green's function formalism in BKMH-$J$ model, we also calculate $C_{\uparrow}$ inside the inter-layer dimer-singlet insulator phase at finite $U$. In the $U$-$J$ phase diagram presented in Fig.~\ref{fig:BKMHLatticeGeometry} (b), we choose a path at $\lambda=0.2t,U=t$ with $J\in[3.3,5.0]$ (the magenta path in Fig.~\ref{fig:BKMHLatticeGeometry} (b)). The Chern number $C_{\uparrow}$ calculated directly from $L=12$ QMC simulation and after the periodization scheme are presented in Fig.~\ref{fig:BKMHChernU1Period}. Again, for the finite-$U$ region inside the inter-layer dimer-singlet insulator phase, spin Chern number $C_s$ defined from single-particle Green's function possesses a $C_s=+2$ value, which is indeed unexpected. Regardless of $U=0$ or $U\neq0$, the inter-layer dimer-singlet insulator phase should be topologically trivial (which has been confirmed by the absence of edge states from strange correlator measurements in QMC~\cite{He2015a}). So now, it is clear that the spin Chern number can not correctly describe this $J$-driven TPT in BKMH-$J$ model.

The inter-layer dimer-singlet insulator phase in BKMH-$J$ model is a Mott insulator without any spontaneous symmetry breaking, and the $J$-driven TPT is TI-to-Mott-insulator transition. Furthermore, the inter-layer dimer-singlet insulator phase cannot be adiabatically connected to any noninteracting band insulator without phase transition and symmetry breaking. This is in sharp contrast with the cases in both Sec.~\ref{sec:GKMH} and Sec.~\ref{sec:CKMH}, where the dimer insulator phase in GKMH model and pVBS, cVBS phases in CKMH model after the $U$-driven TPTs have their respective noninteracting correspondences. At first glance, one might think that at the $J\to\infty$ limit, the $J$-singlet phase is also adiabatically connected to a noninteracting insulator with large inter-layer hopping $t_z$. Surely, a large $t_z$ can give rise to a trivial band insulator. However, these two phases actually have different charge symmetries, which are $U(1)\times U(1)$ for $J$-singlet phase and only a single $U(1)$ for large-$t_z$ insulator phase respectively. According to this analysis, the nonexistent adiabatic connection to a band insulator is the essential reason for the breakdown of spin Chern number to capture the inter-layer dimer-singlet insulator phase in BKMH-$J$ model.

\subsection{Interaction-driven TPT in BKMH-$V$ model}
\label{sec:BKMHV}

The BKMH-$J$ model is not a special case for the breakdown of the Green's function formalism of topological invariants. As proposed in Ref.~\cite{You2015}, there are a series of interaction-driven TPT in 2D interacting TIs, where the single-particle gap does not close at the transition, and there is no noninteracting correspondence for both the TPT itself and the topological trivial insulator phase after the transition. The bilayer Kane-Mele-Hubbard model with inter-layer interaction $V$ (BKMH-$V$ model) in this section, is the second explicit example.

\begin{figure}[t]
\centering
\includegraphics[width=\columnwidth]{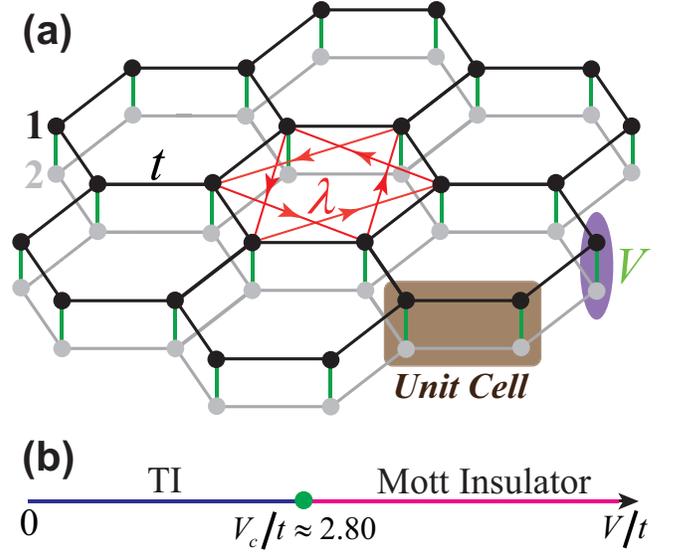}
\caption{\label{fig:BKMHLatticeV} (Color online) (a) Illustration of BKMH-$V$ model. All the terms in the Eq.~\ref{eq:ModelHamiltonian-BKMH-V} are identical to those in Eq.~\ref{eq:ModelHamiltonian-BKMH-J} except the inter-layer AFM interaction $J$ is replaced by the inter-layer electron repulsion $V$. (b) The phase diagram of BKMH-$V$ model. $V$-driven TPT from 2D TI to the featureless Mott insulator is at $V_c/t\simeq 2.80$. }
\end{figure}

The Hamiltonian is given as
\begin{eqnarray}
H =&& -t\sum_{\xi\langle i,j
\rangle,\alpha}( c^{\dagger}_{\xi i\alpha}c_{\xi j\alpha} + c^{\dagger}_{\xi j\alpha}c_{\xi i\alpha} ) \nonumber \\
  &&+ i\lambda\sum_{\xi\langle\!\langle i,j \rangle\!\rangle, \alpha\beta}
v_{ij}(c^{\dagger}_{\xi i\alpha}\sigma^{z}_{\alpha\beta}c_{\xi j\beta} - c^{\dagger}_{\xi j\beta}\sigma^{z}_{\beta\alpha}c_{\xi i\alpha}) \nonumber \\
  &&+ V\sum_{i}( c_{1i\uparrow}^{\dagger}c_{2i\uparrow}c_{1i\downarrow}^{\dagger}c_{2i\downarrow}
     + c_{2i\downarrow}^{\dagger}c_{1i\downarrow}c_{2i\uparrow}^{\dagger}c_{1i\uparrow} ),
 \label{eq:ModelHamiltonian-BKMH-V}
\end{eqnarray}
where both the lattice geometry and the model parameters are depicted in Fig.~\ref{fig:BKMHLatticeV} (a). At first glance, this model preserves $U(1)_{\text{charge}}\times[U(1)\times U(1)]_{\text{spin}}\rtimes Z_2^T$ symmetry. Here the two $U(1)$ spin symmetries correspond to the $S_z$ conservations in each layer. A more careful analysis reveals that this model also preserves the $SO(4)\simeq SU(2)\times SU(2)$ symmetry similar to the BKMH-$J$ model at $U=0$ in Sec.~\ref{sec:BKMHJ}, since the model Hamiltonian in Eq.~\ref{eq:ModelHamiltonian-BKMH-V} can also be written by the $f$-fermions as
\begin{eqnarray}
\label{eq:BKMHQMCFfermionV}
H = \sum_{i,j,\sigma}\chi_{\sigma}(f_{i\sigma}^{\dagger}t_{ij}f_{j\sigma} + h.c.)
              - \frac{V}{2}\sum_{i}(\hat{P}_{i}\hat{P}_{i}+ \hat{P}_{i}^{\dagger}\hat{P}^{\dagger}_{i}), \hspace{0.5cm}
\end{eqnarray}
where $\chi_{\sigma}$, $t_{ij}$, and $\hat{P}_{i}$ are all the same as those in Eq.~\ref{eq:BKMHQMCFfermion}. This Hamiltonian is also invariant under the transformation as $f_{i\sigma}\to U_{\sigma}f_{i\sigma}$ with $U_{\sigma}\in SU(2)$ for $\sigma=\uparrow,\downarrow$, independently, so the BKHM-$V$ model indeed has the $SO(4)\simeq SU(2)\times SU(2)$ symmetry.

At $V\to\infty$ limit, one can actually obtain the exact many-body ground state wavefunction of Eq.~\ref{eq:BKMHQMCFfermionV}, which is a direct product state as
\begin{eqnarray}
\label{eq:SO4ModelVge0}
|\Psi_g\rangle = \prod_i|\Psi_i\rangle=\prod_i\frac{1}{\sqrt{2}}(c_{1i\uparrow}^{\dagger}c_{1i\downarrow}^{\dagger}-c_{2i\uparrow}^{\dagger}c_{2i\downarrow}^{\dagger})|0\rangle.
\end{eqnarray}
This ground state $|\Psi_g\rangle$ is indeed a featureless Mott insulator. It does not break the underlying $SO(4)$ symmetry of the Hamiltonian explicitly or spontaneously, as all the bilinear fermion condensations vanish. For example, one can easily verify that $\langle\Psi_g|c_{\xi i\alpha}^{\dagger}c_{\eta j\beta} |\Psi_g\rangle=\frac{1}{2}\delta_{\xi\eta}\delta_{ij}\delta_{\alpha\beta}$ and $\langle\Psi_g|c_{\xi i\alpha}^{\dagger}c_{\eta j\beta}^{\dagger} |\Psi_g\rangle=\langle\Psi_g|c_{\xi i\alpha}c_{\eta j\beta} |\Psi_g\rangle=0$. What's more, $|\Psi_g\rangle$ does not have noninteracting correspondence either since one can simply observe the double occupancy of $|\Psi_g\rangle$ is $\langle\Psi_g|n_{1i\uparrow}n_{1i\downarrow}|\Psi_g\rangle=\langle\Psi_g|n_{2i\uparrow}n_{2i\downarrow}|\Psi_g\rangle=\frac{1}{2}$, whereas the double occupancy for a noninteracting system is $\frac{1}{4}$. Hence, the ground state $|\Psi_g\rangle$ in Eq.~\ref{eq:SO4ModelVge0} for the model in Eq.~\ref{eq:ModelHamiltonian-BKMH-V} under $V\to+\infty$ limit is a featureless Mott insulator.

\begin{figure}[tp!]
\centering
\includegraphics[width=\columnwidth]{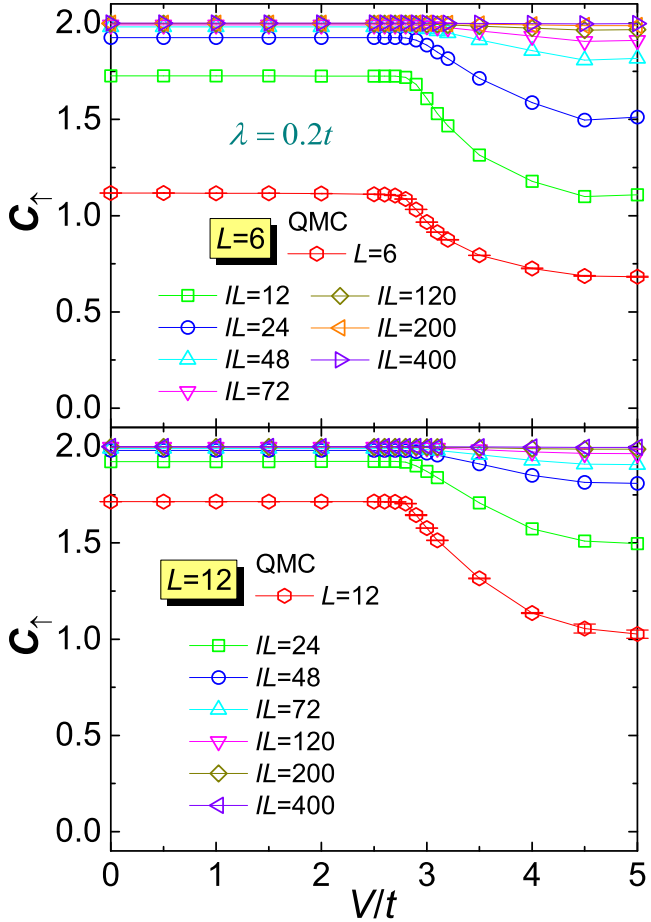}
\caption{\label{fig:BKMHVChernPeriod}(Color online) Chern number $C_{\uparrow}$ for the $V$-driven TPT in BKMH-$V$ model with $\lambda=0.2t$  from QMC and periodization results using the QMC data in (a) $L=6$ and (b) $L=12$ systems. The ideally quantized Chern number $C_{\uparrow}$ in large $IL$ cases indicate that $C_{\uparrow}$ has no variation across this topological phase transition.}
\end{figure}

Our QMC results of BKMH-$V$ model reveal that there is a $V$-driven TPT from TI to Mott insulator at $V_c/t\simeq 2.82$. The corresponding phase diagram is presented in Fig.~\ref{fig:BKMHLatticeV} (b). We also confirm that the single-particle gap does not close across this TPT. Instead, the charge $2e$ excitation gap, corresponding to the on-site spin-singlet $s$-wave pairing order $\hat{\Delta}_i^{\dagger}=\frac{1}{\sqrt{2}}(c_{1i\uparrow}^{\dagger}c_{1i\downarrow}^{\dagger}-c_{2i\uparrow}^{\dagger}c_{2i\downarrow}^{\dagger})$, closes at the transition point. Both the single-particle gap and the on-site pairing gap across the topological phase transition for the BKMH-$V$ model are presented in Appendix~\ref{sec:appendix_a}. After the transition, the system enters into a featureless Mott insulating phase.

To demonstrate the breakdown of the Green's function formalism across this $V$-driven TPT, we calculate the spin Chern number $C_s$ for BKMH-$V$ model. The results of $C_{\uparrow}$ from QMC simulations of $L=6,12$ are presented in Fig.~\ref{fig:BKMHVChernPeriod} (a) and Fig.~\ref{fig:BKMHVChernPeriod} (b). Again, we can observe that $C_{\uparrow}$ varies continuously without finite-value jump for both $L=6$ and $L=12$ systems. By further applying the periodization scheme to the data of $L=6,12$ systems, we obtain the integer-quantized Chern number $C_{\uparrow}$, also presented in Fig.~\ref{fig:BKMHVChernPeriod} (a) and Fig.~\ref{fig:BKMHVChernPeriod} (b). Same as the one in the BKMH-$J$ model, the Chern number $C_{\uparrow}$ acquires no change across the $V$-driven TPT, since the $SO(4)$ $V$-singlet insulator at $V>V_c$ is a product state with all the electron degrees of freedom frozen, i.e., no edge states. The ideally quantized spin Chern number inside the featureless Mott insulator phase is another manifestation of the failure of the topological invariants constructed from the Green's function formalism.

\section{How the spin Chern number works?}
\label{sec:TopoCondition}

Based on the above QMC results in Sec.~\ref{sec:GKMH}, ~\ref{sec:CKMH}, ~\ref{sec:BKMHJ}, and ~\ref{sec:BKMHV} on interaction-driven topological phase transitions, we can now arrive at some basic understanding on the reason why the spin Chern number constructed from single-particle Green's function works well in some interacting topological quantum phases while it experiences breakdown in the others.

\subsection{Working condition of the spin Chern number}
\label{sec:Working}
In the interaction-driven TPTs in GKMH (Sec.~\ref{sec:GKMH}) and CKMH (Sec.~\ref{sec:CKMH}) models, all the phases can be adiabatically connected to the corresponding noninteracting insulator phases in the same model Hamiltonian, and at the transitions, the single-particle gap closes, just like in the corresponding noninteracting cases. In these cases, the spin Chern number can be successfully applied to characterize topologically distinct phases and the TPTs.
The underlying physics goes as follows. For the spin Chern number $C_s$ constructed from the single-particle Green's function to acquire a change of integer number, the zero-frequency single-particle Green's function must possess either a pole or a zero~\cite{He2015b}. The appearance of a pole corresponds to the single-particle gap closing, as observed in all the topological phase transitions in free fermion systems. The appearance of a zero in the single-particle Green's function across TPT has been discussed and found in interacting TIs~\cite{Yoshida2014,You2014b,Slagle2015}. Here, we have confirmed that the single-particle gap close at the $U$-driven TPTs in both GKMH and CKMH models. As for the $J$-driven TPT in BKMH-$J$ and $V$-driven TPT in BKMH-$V$ models, we find that across the transition, the single-particle gap keeps finite and there is neither pole or zero appearing in the zero-frequency single-particle Green's function.

For a noninteracting TI with finite spin Chern number $C_s$, its spin Hall conductivity is $\sigma_{xy}^{spin}=C_s\frac{e^2}{h}$. And for the free fermion system, the spin Chern number can be calculated from the single-particle Green's function~\cite{Avron1983,So1985,Volovik1988,volovik2009universe}. For an interacting TI with finite $C_s$, if it is adiabatically connected to a noninteracting TI without phase transition and symmetry breaking, then this interacting TI should have exactly the same physical spin Hall conductivity. So we can conclude that in interacting systems if both phases across a TPT (driven by interaction or one-body-parameter and without explicit or spontaneous symmetry breaking) can be adiabatically connected to their noninteracting correspondences, then the spin Chern number calculated from single-particle Green's function can always characterize them and the transition.

\subsection{Breakdown of the spin Chern number constructed from the Green's function}
\label{sec:Breakdown}

The reason for the breakdown of the spin Chern number constructed from the Green's function in the BKMH-$J$ (Sec.~\ref{sec:BKMHJ}) and BKMH-$V$ (Sec.~\ref{sec:BKMHV}) models is two-folded. First, the inter-layer dimer-singlet insulator in BKMH-$J$ model and the featureless Mott insulator in BKMH-$V$ model after the transitions are Mott insulators without noninteracting correspondence. Second, the critical fluctuation associated to the transition is collective and bosonic instead of single-particle and fermionic.

Across the transition, neither pole nor zero of single-particle Green's function appears, which results in the same integer values of the spin Chern number.
In the inter-layer dimer-singlet insulator and featureless Mott insulator phase, as shown in Fig.~\ref{fig:BKMHChernU0Period}, Fig.~\ref{fig:BKMHChernU1Period}, and Fig.~\ref{fig:BKMHVChernPeriod}, the spin Chern number constructed from the single-particle Green's function is equal to $C_s=+2$, which is an artifact of the Green's function formalism.

We can understand such artifact from the perspective of symmetry-protected topological (SPT) phases~\cite{Chen2012,Chen2013}. An important property of SPTs is that they only have short-range entanglement and can be adiabatically connected to some direct product state (with the same topological invariant) without going through symmetry breaking and phase transitions. The simplest product state for noninteracting fermion systems is the Slater determinant, i.e., a product state of free fermion wavefunction in momentum space. The noninteracting correspondences of the trivial insulators after the $U$-driven TPTs in GKMH and CKMH models are such Slater determinants. However, for BKMH-$J$ and BKMH-$V$ models, although the inter-layer dimer-singlet insulator and featureless Mott insulator at large $J$ and $V$ can be adiabatically connected to the product states at the limit of $J\to+\infty$ and $V\to+\infty$, the wavefunction basis of such product states are singlets consisting of two electrons, instead of the single-electron wavefunction used to construct the Slater determinant. Thus, the inter-layer dimer-singlet insulator and featureless Mott insulator in BKMH-$J$ and BKMH-$V$ models are not adiabatically connected to free-fermion Slater determinants. In the cases of BKMH-$J$ and BKMH-$V$ models, even if one constructs the topological invariants from single-particle Green's function formalism, the obtained spin Chern numbers do not correspond to the physical spin Hall conductivity. The physical spin Hall conductivity, on the other hand, should be carried by the emergent low-energy bosonic modes, which become critical at these interaction-driven TPTs~\cite{He2015a,You2015}.

\section{Summary}
\label{sec:Summary}

By means of large-scale QMC simulations, we have investigated several interaction-driven topological phase transitions in 2D interacting TIs without explicit or spontaneous symmetry breaking. We further characterize these TPTs via the topological invariants across these interaction-driven TPTs, including $Z_2$ invariant and spin Chern number constructed from single-particle Green's function at zero frequency. We find that the spin Chern number successfully detects the interaction-driven TPTs in GKMH (Sec.~\ref{sec:GKMH}) and CKMH (Sec.~\ref{sec:CKMH}) models, while it surprisingly experiences breakdown in BKMH-$J$ (Sec.~\ref{sec:BKMHJ}) and BKMH-$V$ (Sec.~\ref{sec:BKMHV}) models. To understand such breakdown, we have analyzed the working condition for spin Chern number constructed in Green's function formalism, and discuss why it fails for the inter-layer dimer-singlet insulator and featureless Mott insulator phases in BKMH-$J$ and BKMH-$V$ models. It turns out that the spin Chern number constructed from single-particle Green's function can only characterize interacting TIs which can be adiabatically connected to noninteracting insulators, for which the spin Chern number corresponds to the physical spin Hall conductivity. For interacting TIs without noninteracting correspondence, the spin Chern number constructed from single-particle Green's function is artificial, as demonstrated in BKMH-$J$ and BKMH-$V$ models.

In terms of SPT, the inter-layer dimer-singlet insulator and the featureless Mott insulator in BKMH-$J$ and BKMH-$V$ models, where the spin Chern number experiences breakdown, are indeed SPT trivial states. Actually, the interaction-driven TPTs in these models are of bosonic nature due to the gapped fermion degree of freedom. Our work highlights the important issue of how to characterize the topological aspects of SPT states in generally interacting fermion systems which cannot be adiabatically connected a noninteracting band insulator. As a result, we expect new and more versatile techniques to correctly describe the topological invariants in such interacting states. Recent progress in calculating entanglement spectrum~\cite{Assaad2014,Assaad2015}, entanglement entropy~\cite{DaWang2015}, and directly probing the edge states via strange correlation~\cite{You2014_SC,Wu2015} in interacting TIs seem to provide promising directions.

\begin{acknowledgments}
We thanks Yi-Zhuang You, Lei Wang, Ning-Hua Tong, Zheng-Xin Liu, Zhong Wang, Cenke Xu, Liang Fu, Kai Sun, Xi Dai for inspiring discussions on various aspects of this paper. The numerical calculations were carried out at the Physical Laboratory of High Performance Computing in Renmin University of China, the supercomputing platforms in the Center for Quantum Simulation Sciences in the Institute of Physics, Chinese Academy of Sciences, as well as the National Supercomputer Center in TianJin on the platform Tianhe-1A. YYH, HQW and ZYL acknowledge support from National Natural Science Foundation of China (Grant Nos.91421304 and 11474356). ZYM is supported by the National Natural Science Foundation of China (Grant Nos.11421092 and 11574359) and the National Thousand-Young-Talents Program of China, and acknowledges the hospitality of the KITP at the University of California, Santa Barbara, where part of the this work is initiated.
\end{acknowledgments}

\appendix

\section{Excitation gaps across the TPT in BKMH-$V$ model}
\label{sec:appendix_a}

For BKMH-$V$ model in Eq.~\ref{eq:ModelHamiltonian-BKMH-V}, we have mentioned that the single-particle gap always keeps finite while the charge gap (on-site pairing gap) closes and reopens across the topological phase transition, around $V_c/t=2.80$. Here, we present numerical data of both gaps for BKMH-$V$ model.
\begin{figure}[t]
\centering
\includegraphics[width=\columnwidth]{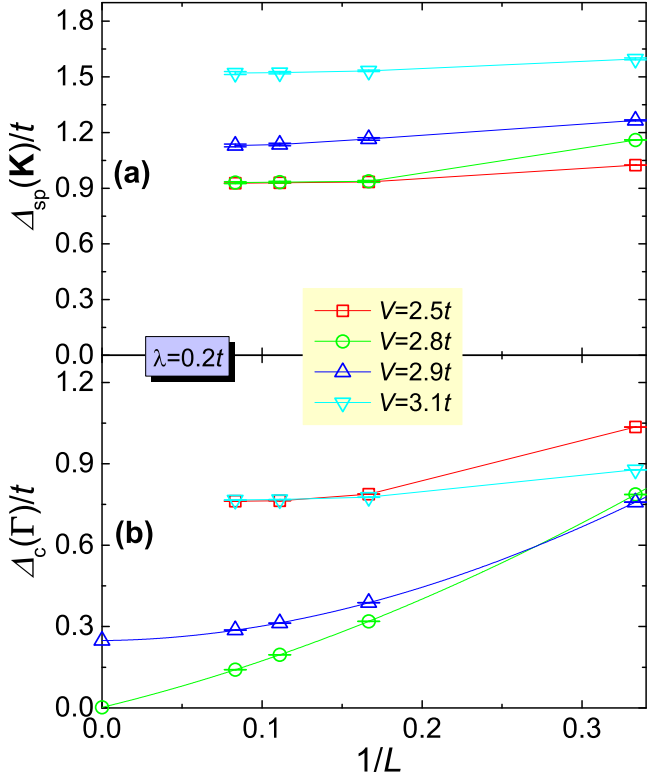}
\caption{\label{fig:BKMHVGaps}(Color online) (a) Single-particle gap $\Delta_{sp}(\mathbf{K})$ and (b) charge gap $\Delta_c(\mathbf{\Gamma})$ for BKMH-$V$ model with $\lambda=0.2t$ for chosen $V/t=2.5,2.8,2.9,3.1$. $\Delta_{sp}(\mathbf{K})$ keeps finite across the TPT, while $\Delta_c(\mathbf{\Gamma})$ experiences a closing and reopening at $V_c/t\simeq 2.80$ after the extrapolation to thermodynamic with third-order polynomial in $1/L$. }
\end{figure}

The single-particle gap is simply extracted from the imaginary-time single-particle Green's function in reciprocal space as $G_{\sigma}(\tau,\mathbf{k})$, which is a $4\times 4$ Hermitian matrix for BKMH-$V$ model. At the $\tau\to+\infty$ limit, we have $[G_{\sigma}(\tau,\mathbf{k})]_{\alpha\alpha}\propto Z_{\mathbf{k}}e^{-\tau\Delta_{sp}(\mathbf{k})}$ with $\alpha=1,2,3,4$ indicating sublattices, and $\Delta_{sp}(\mathbf{k})$ is the single-particle gap at $\mathbf{k}$ point for the system. As for the on-site pairing gap, we define the spin-singlet $s$-wave pairing order
\begin{eqnarray}
\label{eq:PairingOrder}
\hat{\Delta}_{i\alpha}^{\dagger}=\frac{1}{\sqrt{2}}(c_{1i\alpha\uparrow}^{\dagger}c_{1i\alpha\downarrow}^{\dagger}
                                -c_{2i\alpha\uparrow}^{\dagger}c_{2i\alpha\downarrow}^{\dagger}),
\end{eqnarray}
where $\alpha=1,2$ stands for $A, B$ sublattices respectively and integer $i$ represents the unit cells of the bilayer model. We can observe that $\hat{\Delta}_{i\alpha}^{\dagger}$ represents certain kind of local pairing order on vertical bonds between layers. Then we can obtain the dynamic correlation function in reciprocal space for such on-site pairing order as
\begin{eqnarray}
\label{eq:PairingOrderCorrelation}
[P(\mathbf{Q},\tau)]_{\alpha\beta}=\frac{1}{N}\sum_{ij}e^{i\mathbf{Q}\cdot(\mathbf{R}_i-\mathbf{R}_j)}\langle T_{\tau}[ \hat{\Delta}_{i\alpha}^{\dagger}(\tau)\hat{\Delta}_{j\beta}(0) ]\rangle, \hspace{0.4cm} \nonumber \\
\end{eqnarray}
where $\mathbf{Q}=\mathbf{\Gamma}$ is the ordering vector for BKMH-$V$ model and $N=L^2$ is the number of unit cells for a $L\times L$ system. Then we can extract the corresponding charge gap $\Delta_c(\mathbf{Q})$ under $\tau\to+\infty$ limit via $[P(\mathbf{Q},\tau)]_{\alpha\alpha}\propto R_{\mathbf{Q}}e^{-\tau\Delta_c(\mathbf{Q})}$.

Both the numerical results of single-particle gap $\Delta_{sp}(\mathbf{K})$ and charge gap $\Delta_c(\mathbf{\Gamma})$ across the TPT in BKMH-$V$ model with $\lambda=0.2t$ are shown in Fig.~\ref{fig:BKMHVGaps}. The single-particle gap $\Delta_{sp}(\mathbf{K})$ (Fig.~\ref{fig:BKMHVGaps} (a)) keeps finite with a quite large value, which is about $0.9t$, at the topological phase transition point. However, the charge gap $\Delta_c(\mathbf{\Gamma})$ acquires a closing and reopening across the TPT around $V_c/t\simeq 2.80$.

\begin{figure}[h!]
\centering
\includegraphics[width=\columnwidth]{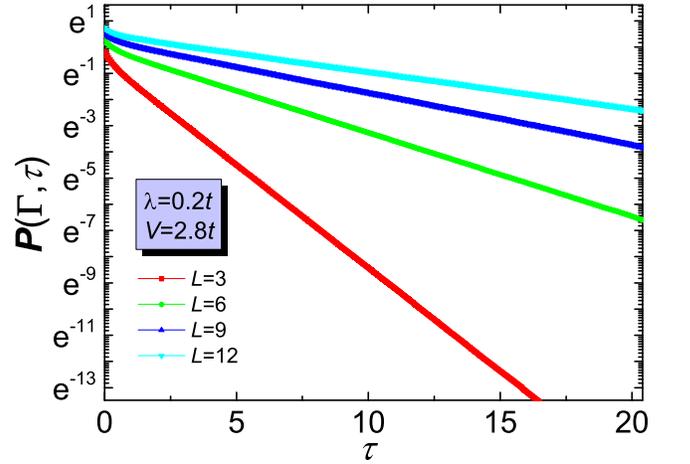}
\caption{\label{fig:OriginalData}(Color online) Original data of $[P(\mathbf{Q},\tau)]_{11}$ at topological phase transition point $V/t=2.80$ for BKMH-$V$ model systems with $L=3,6,9,12$, under semilogarithmic plot. }
\end{figure}

We emphasize that the original data of $[P(\mathbf{Q},\tau)]_{11}$, via which we extract the charge gap $\Delta_c(\mathbf{\Gamma})$, has quite high quality. For example, we present $[P(\mathbf{Q},\tau)]_{11}$ data for $V/t=2.80$ in BKMH-$V$ model systems with $L=3,6,9,12$ in Fig.~\ref{fig:OriginalData} with semilogarithmic plot. In Fig.~\ref{fig:OriginalData}, we can see that the data of $\ln[P(\mathbf{Q},\tau)]_{11}$ is perfectly linear with imaginary-time $\tau$. Thus, the closing of charge gap $\Delta_c(\mathbf{\Gamma})$ around $V_c/t\simeq 2.80$ at the $V$-driven topological phase transition point in BKMH-$V$ model, as shown in Fig.~\ref{fig:BKMHVGaps} (b), is sufficiently solid.

\bibliography{SpinChernII_Bib}

\end{document}